\documentclass{llncs}
\usepackage{booktabs} 
\usepackage{comment}
\usepackage{graphicx}
\usepackage{cellspace}
\usepackage{wrapfig,picins}
\usepackage[sorting=none]{biblatex}
\usepackage{amsmath}
\usepackage{graphicx}
\usepackage{color}
\usepackage{url}
\usepackage{booktabs} 
\usepackage{subfigure}
\usepackage{amsmath}
\DeclareGraphicsExtensions{.pdf,.png,.jpg}
\usepackage[english]{babel}

\usepackage[section]{placeins}

\usepackage{listings}
\usepackage{xcolor}
\definecolor{codegreen}{rgb}{0,0.6,0}
\definecolor{codegray}{rgb}{0.5,0.5,0.5}
\definecolor{codepurple}{rgb}{0.58,0,0.82}
\definecolor{backcolour}{rgb}{0.95,0.95,0.92}
\lstdefinestyle{mystyle}{
    backgroundcolor=\color{backcolour},   
    commentstyle=\color{codegreen},
    keywordstyle=\color{magenta},
    numberstyle=\tiny\color{codegray},
    stringstyle=\color{codepurple},
    basicstyle=\ttfamily\footnotesize,
    breakatwhitespace=false,         
    breaklines=true,                 
    captionpos=b,                    
    keepspaces=true,                 
    numbers=left,                    
    numbersep=5pt,                  
    showspaces=false,                
    showstringspaces=false,
    showtabs=false,                  
    tabsize=2
}
\begin{document}

\frontmatter          
\pagestyle{empty}  

\title{Teaching Math with the help of Virtual Reality}
%
%
 \author{Marco Simonetti\inst{1} $^{ORCID: 0000-0003-2923-5519}$
 \newline Damiano Perri\inst{1} $^{ORCID: 0000-0001-6815-6659}$
 \newline Natale Amato\inst{2} $^{ORCID: 0000-0001-5104-9494}$
 \newline Osvaldo Gervasi\inst{3} $^{ORCID: 0000-0003-4327-520X}$ 
 }
\institute{
University of Florence, Dept. of Mathematics and Computer Science, Florence, Italy \and University of Bari, Bari (Italy)\and University of Perugia, Dept. of Mathematics and Computer Science, Perugia, Italy
}
\titlerunning{Teaching Math with the help of Virtual and Augmented Reality} 
\authorrunning{Damiano Perri, Marco  Simonetti, Amato Natale and Osvaldo Gervasi} 

\maketitle

\begin{abstract}
In the present work we intend to introduce a system based on VR (Virtual Reality) for examining analytical-geometric structures that occur in the study of mathematics and physics concepts in the last high school classes.\newline
In our opinion, an immersive study environment has several advantages over traditional two-dimensional environments (such as a book or the simple screen of a PC or tablet), such as the spatial understanding of the concepts exposed, more peripheral awareness and moreover an evident decreasing in the information dispersion phenomenon. This does not mean that our pedagogical approach is a substitute for traditional pedagogical approaches, but is simply meant to be a robust support.
In the first phase of our research we have tried to understand which mathematical objects and which tools to use to enhance mathematical teaching, to demonstrate that the use of VR techniques significantly increase the level of understanding of the mathematical subject investigated by the students.\newline
The system which provides for the integration of two machine levels, hardware and software, was subsequently tested by a representative sample of students who returned various food for thought through a questionnaire.\newline

\end{abstract}


\section{Introduction}
Much progress has been made in the field of VR since it was introduced in the early 70s of the last century and in many fields it has found innumerable applications such as entertainment\cite{Gervasi2009}, teaching\cite{vr2018Casiduso}, tourism\cite{Williams1995}\cite{YoungRyan2019}, manufacturing industry\cite{Doil2003}\cite{Mujber2004}, networking and communications\cite{Lazar1992}\cite{Russel2001}, microelectronic and hi-performances hardware industries\cite{VellaGPU}\cite{VellaNGT12}, e-commerce\cite{Papadopoulou2007}, medicine\cite{vrrehabilitation}\cite{Sheng2020}.\newline
Today, we have reached a certain maturity of VR techniques and in this work we intend to deepen some themes concerning simple VR and even AR (Augmented Reality) experiences in the teaching of mathematics.\newline
During our work, we were struck several times by a precise thought: the possibility of giving real form to the ideal objects of mathematics has always been a dream caressed by the mind of man who felt the need to convey his ideas to others without losing their purity and meaningfulness.\newline From the primitive cave representations of simple arithmetic objects, to the elegant structures of the symbolic algebra of the seventeenth century, through the imposing constructions of mathematical analysis and modern geometry, to get to the current and amazing views of numerical analysis through computer graphics.
So, we set out to investigate the possibility of extending students' understanding of the concept of link between an algebraic-set structure and its geometric representation on an orthogonal Cartesian space (function).\newline
Obviously, in the first phase of the work, the attention was focused on a limited number of functions, significant for the students of the terminal classes, some representing trajectories in the Cartesian plane, others as surfaces in the three-dimensional space.\newline
The very first proposed functions are as follows:

\begin{itemize}
    \item $y = \sin x$
    \begin{description}
    \item
    The \textbf{sine function} is well known by students who are currently using it to solve analytical and geometric problems, as well as is commonly used in physics to model periodic phenomena such as sound and light waves or even changes in average temperature during the day or the year\cite{Boyer2011}.
    As showed in fig.\ref{fig:sinewave}
    \end{description}
    \begin{figure}[h]
        \centering
        \includegraphics[width=0.5\textwidth]{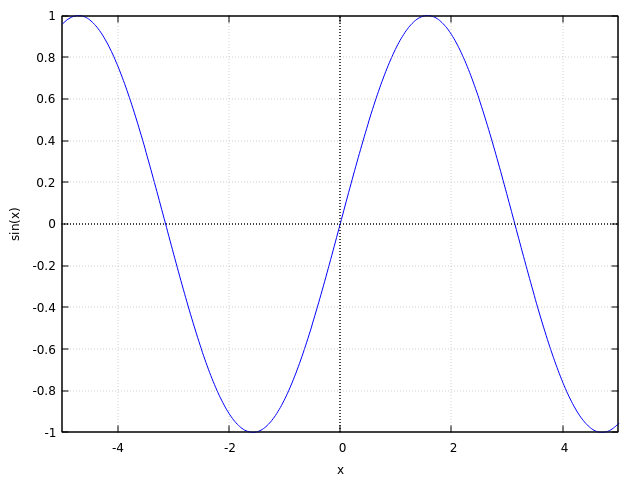}
        \caption{sinewave}
        \label{fig:sinewave}
    \end{figure}
    
    \item $z = x^2-y^2$
    \begin{description}
    
      \item 
    This function is a classical example of a \textbf{saddle surface} that it is a smooth surface containing one or more saddle points.\newline
    Saddle surfaces have negative Gaussian curvature which distinguish them from convex/elliptical surfaces which have positive Gaussian curvature and are very important in the study of non-Euclidean geometry and in the theory of general relativity\cite{Creighton2003}.
    As showed in fig.\ref{fig:x2-y2}
    \end{description}
    
    \item $z = xy$
    \begin{description}
    
      \item 
        This function is an example of \textbf{hyperbolic paraboloid}\cite{Creighton2003}.
        As showed in fig.\ref{fig:x2-y2}
        \begin{figure}[h]
        \centering
        \includegraphics[width=0.8\textwidth]{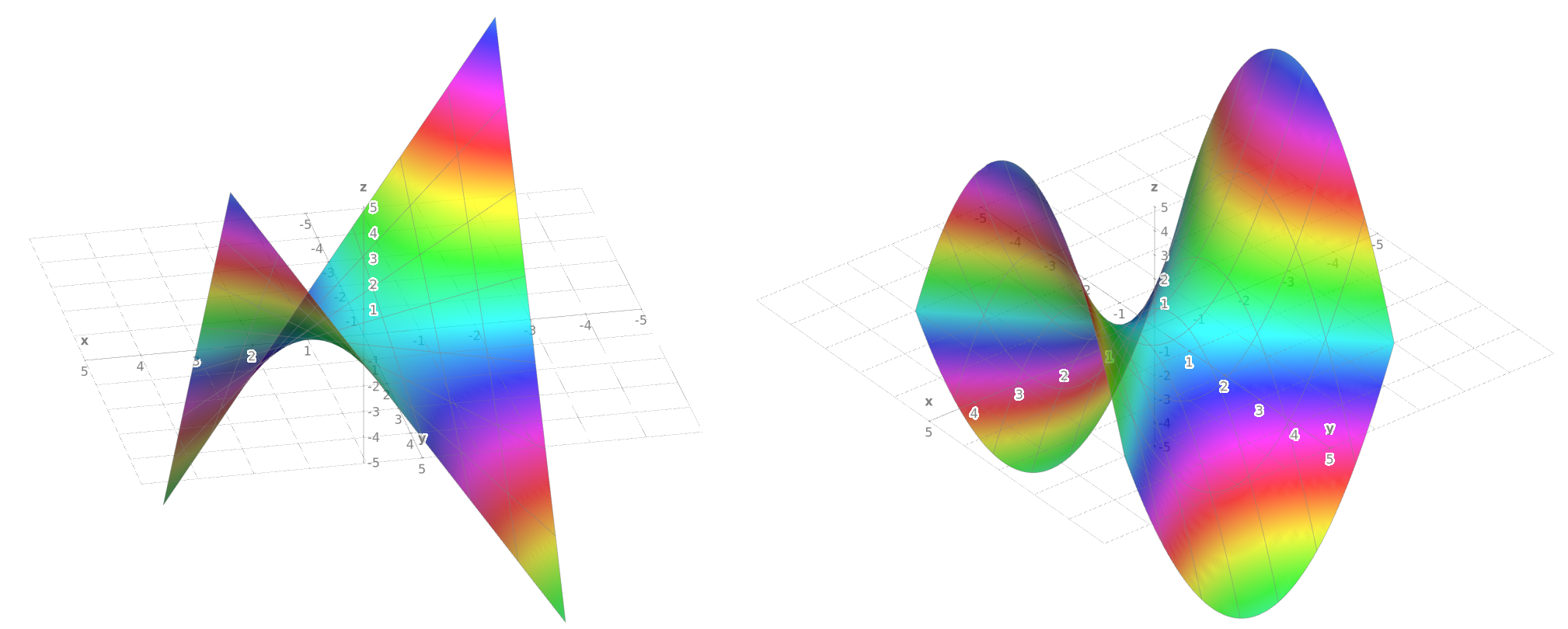}
        \caption{Saddle Surface (\emph{left}) and Hyperbolic Paraboloid (\emph{right})\newline Real and Imaginary part respectively of the function $f(z) = z^2$}
        \label{fig:x2-y2}
        \end{figure}
    \end{description}

    \item $z = \ln (x^2+y^2)$
    \begin{description}
      \item 
        This function is an example of \textbf{bi-dimensional logarithm}, useful to describe astrophysical objects.
        As showed in fig.\ref{fig:ln}
        \begin{figure}[h]
        \centering
        \includegraphics[width=0.6\textwidth]{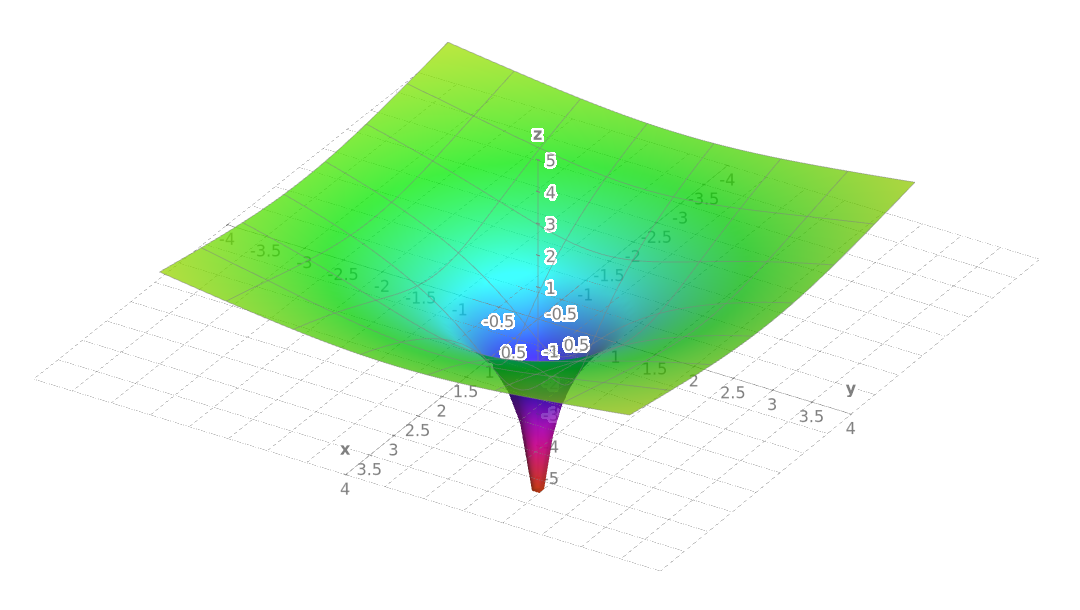}
        \caption{Bi-dimensional Natural Logarithm}
        \label{fig:ln}
    \end{figure}
    \end{description}

    \item $z = \displaystyle \frac{\sin (x^2+y^2)}{x^2+y^2}$
    \begin{description}
      \item This function is an example of \textbf{bi-dimensional dumped sine}, useful to describe objects in fluid dynamics, electronics and telecommunications.
        As showed in fig.\ref{fig:sin}
        \begin{figure}[h]
        \centering
        \includegraphics[width=0.6\textwidth]{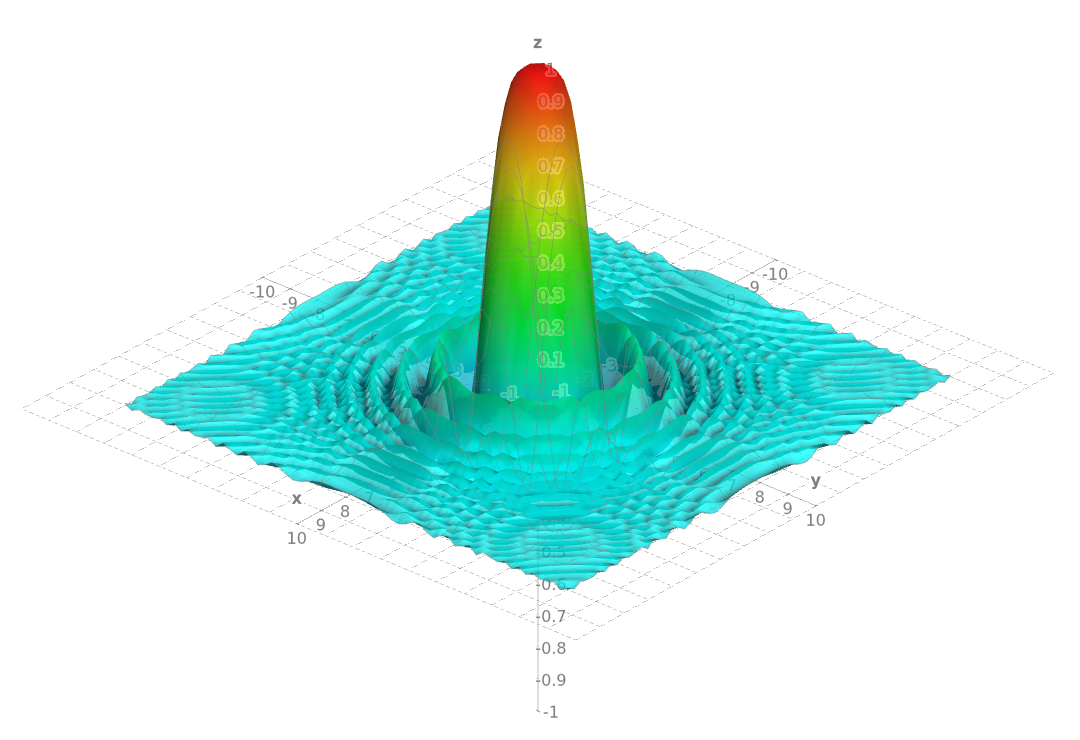}
        \caption{Dumped Sine}
        \label{fig:sin}
    \end{figure}
    \end{description}

\end{itemize}
\section{Related works}
Stimulating multiple sensory dimensions during the learning process of a concept or idea makes the result particularly incisive; our mind needs to experience the reality that surrounds it through multiple experiential levels: looking, listening, touching allow us to build a complex conceptual structure that founds and strengthens our knowledge\cite{meehan2002physiological}\cite{raja2004exploring}.\newline
In our opinion, an immersive study environment has several advantages over traditional two-dimensional environments (such as a book or the simple screen of a PC or tablet), such as the spatial understanding of the concepts exposed, more peripheral awareness or more useful information bandwidth and decreasing in the information dispersion phenomenon\cite{bowman2007virtual}.
Several works have been proposed to achieve the purpose, especially in the natural sciences: chemistry\cite{georgiou2007virtual}, biology\cite{tan2013use}, physiology\cite{ryan2004virtual}, physics\cite{savage2010teaching}.
\newline The idea may therefore be to create digital environments to enhance skills, knowledge and competences in math, avoiding anxiety and bad results\cite{Hyesang2016}.\newline
Mathematics can be considered one of the most difficult subjects for many students. A recent study asks questions about traditional way of learning and recommends more active and attractive learning approaches\cite{Freeman2014}.\newline
Studies have shown that immersion in a digital environment can improve education in several ways, because, as we said before, multiple perspectives are activated\cite{Dede}; this fact has a significant value in all areas of mathematics\cite{pasqualotti2002mat3d}.\newline
In several works it has been seen that despite the mathematical simplicity of the concept of function\cite{akkocc2002simplicity}, many students find it difficult to relate its analytical form to the relative graph, as if the intermediate layers existing between the set concept and the analytic-geometric one prevented a clear understanding of the bond\cite{breidenbach1992development}\cite{akkocc2003function}.\newline
It has recently been proven that the use of software capable of explicitly representing the analytic-geometric link existing in the functions can help students develop a positive attitude towards mathematics itself, in terms of attitude, motivation, interest and competence\cite{king2017using}, but we believe that new researches need to be done to grasp the profound implications that exist.

\section{The Architecture of the system}
We have created two different user experiences.
One uses the VR, while the other uses the AR.
In both cases the graphic engine used is the same, Unity3D.
This software allows the composition of virtual environments starting from basic elements called Assets which the scene is composed with.
It also takes care of rendering, real-time lighting calculation and user interaction management.
The fundamental tools that have been used are the following:
\begin{itemize}
\item the game objects, i.e. the basic elements that make up the scene you want the user to view.
\item the scripts, code files written in C\# language which you can execute predefined tasks by, such as managing the appearance of objects on the displayed scene, or the camera movement as a key on the keyboard gets pressed.
\item the colliders, that prevent intersection or collision between the character user is controlling and the objects in the scene
\end{itemize}

The generation of the shapes is realized with two C\# scripts that allow to model any mathematical function in two or three dimensions.
The first script generates the vertices.
The second script receives in input a list of vertices, and generates a three-dimensional figure.
\\so, let us suppose we want to render a three-dimensional function, for example : $f(z)=x*y$.
As we all know a standard mathematical function like this is defined in the continuum space, so if we wanted it to be represented we should need an infinite number of points: the aim is therefore to make the scenario as plausible as possible in a discreet environment, by appropriately choosing the points to be drawn.
So to do that, it must be chosen a well-defined length along the X, Y and Z axis, and the number of points (i.e. the number of vertices) that compose the graph must be predetermined.
In other words, we need to define a grid of points that will define the level of maximum detail we want to achieve.
In addition, it must be kept in mind that the greater the level of detail, the more calculations the user's device will have to perform to display the object on the screen.
Defining a grid of points is equivalent to defining a sampling rate.
This is the same as when you are processing an electronic signal (e.g. an audio signal) and want to convert from a continuous signal to its discrete representation.
If the number of samples is too low we can in fact obtain Aliasing, obtaining an inaccurate representation of the mathematical function we want to show.
It is possible to run the program already with the code just described.
However, this requires to recalculate all forms at runtime each time.
To improve performance we have therefore saved the forms generated with Unity3D inside the filesystem so that they can directly be reloaded at program start, with no need to recalculate all objects from the beginning.
We then processed the shapes with Blender in order to reduce polygonal complexity without changing their information content.
In other words, the complexity of the figures in terms of vertices has been reduced but an user who observed them would not notice any difference.
This is possible using Blender, a software made to process models and three-dimensional objects.
Finally, we have included them again in the Unity3D project.
\\The VR environment has generated and compiled by WebGL technology.
This means that the application is compatible with all devices (computers or smartphones) on the market since the environment is usable through a web browser.
The graphic quality of the scene adapts according to the computational power of the device, while remaining undemanding in terms of hardware requirements.
The scene can be observed through a virtual reality viewer, such as HTC Vive, or through a normal computer monitor.
The user has the possibility to move around the virtual environment using mouse and keyboard.
Inside the environment are visible three-dimensional geometric shapes that allow to understand some mathematical functions otherwise difficult to draw.
\begin{figure}[h]
    \centering
    \includegraphics[width=1\textwidth]{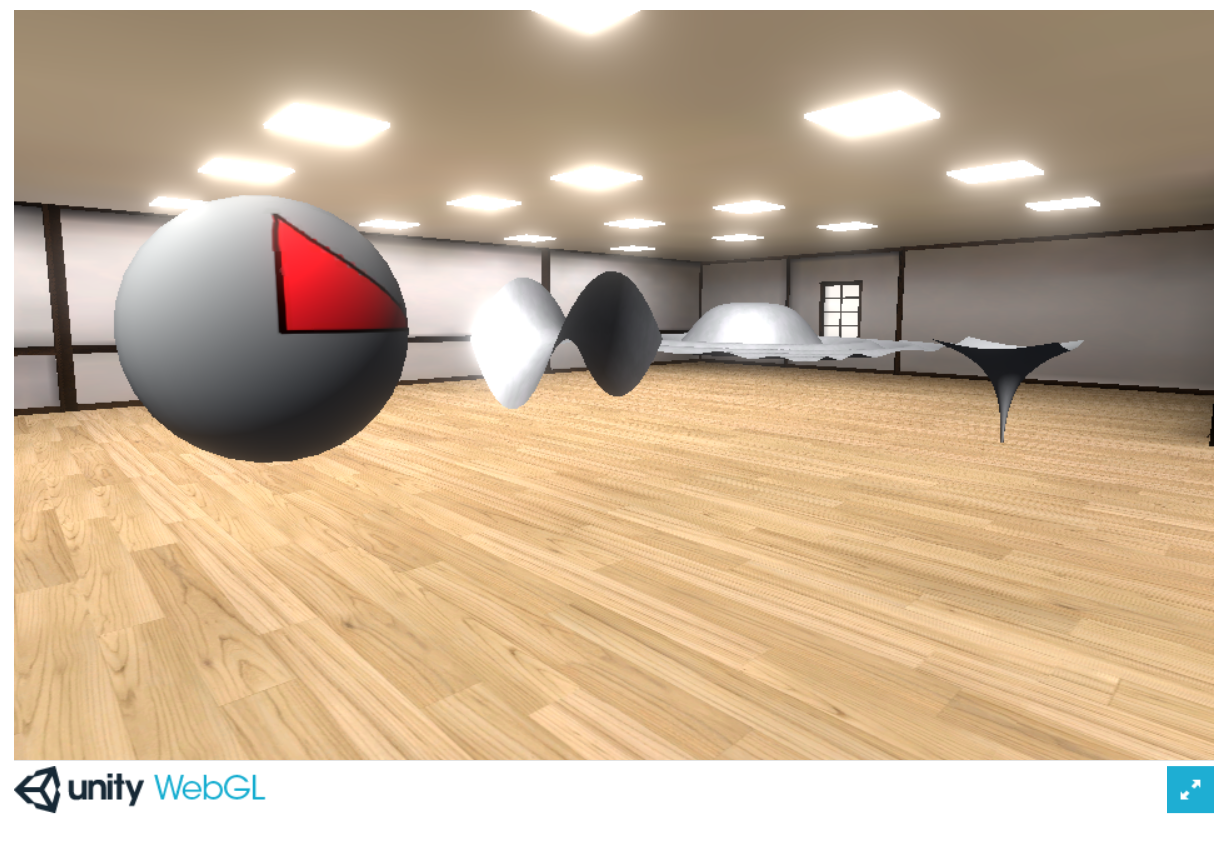}
    \caption{VR Example}
    \label{fig:mesh1}
\end{figure}
The AR environment uses the Vuforia framework.
The program created is an apk, installable on Android smartphones with 7.0+ operating system.
Vuforia is an SDK that allows you to analyze the video stream recorded in real time by the phone camera.
Vuforia allows you to create a database of markers (called Vumark).
These have manually been associated to the game objects of the scene.
When one of the markers present in the database is framed by the camera of the user's device, Vuforia tells Unity to show on the scene (and then on the user's screen) the game object associated with the framed Vumark.
Moreover, this SDK manages the spatial orientation of the object according to the user's position with respect to the Vumark.
If we frame a Vumark and move around it, the object associated with it will rotate as well, allowing us to appreciate it in a realistic way.

\section{The Virtual World made on Unity3D, Blender and Vuforia}
Suppose we want to represent a three-dimensional function within a virtual world created with Unity3D. The creation of the figures is done through two scripts.\newline
First of all it is defined the resolution, that is the level of detail, that the figure must have.\newline
If for example we set a resolution equal to 100, then we will have a matrix of 100x100 points.
In this way we will scroll the variable X and the variable Y along the grid.
Then the step is defined, i.e. how much space must elapse between one point and the next.
The step variable has been fixed at 0.1f.
By means of a nested double "for cycle", where the first takes care of the X variable and the second the Z variable, we can calculate the Y value in the grid.\newline\newline

The code we made allows to obtain the list of points that makes up the figure. The next step is to calculate the list of triangles.
Triangles are a fundamental element in computer graphics.
They allow you to specify how the points are interconnected to each other and how they should be rendered on screen.\newline
Each "game object" and its figure have associated a script of this type.
The difference among the various scripts therefore remains in ability to calculate the getY function.
In fact every time it was necessary to change the graph to be represented, it would be sufficient to specify how to calculate the Y to obtain the change of the shown figure; in all cases the unique thing to be done would be to scan the grid (in our example a 100x100 matrix) and recalculate the correct values.
The fundamental difference is in the case of two-dimensional functions, like the periodic sine function which is mentioned in chapter 2.
In that case only one "for cycle" was sufficient and a depth fixed at 0.05f.
In this way we can represent a two-dimensional function as if it were a tube, which allows us to observe it better when we move around it.\newline
The next step is the generation of the three-dimensional mesh from the vertices and triangles calculated in the previous step.
To do this, a generic script has been created, which can be recalled from all the codes present in the Unity3D project.
Since a three-dimensional object has been generated inside the program, it is necessary to calculate how the light should behave in order to make it visible to the user's camera.\newline
The hard work would therefore seem how to illuminate the object in the right way: that means that it is necessary to calculate a light intensity value for each polygon which makes up the entire object structure.
\newline
Actually, this calculation, which seems very complex, is carried out very quickly in Unity3D.
The calculation is carried out in two phases.
At the beginning, a call to a function integrated in the Unity libraries doubles up triangles in the figure, so each of them has got a "specular twin", with the normal straight line to the surface with opposite direction to its homologous, so that flow of incident light can correctly be obtained.
\newline
Finally, a mesh object is created and vertices, normals and triangles are assigned to it.
At the end of this operation the figure (also called mesh) is ready to be shown on screen.\newline\newline
We have then divided the following work into two different ramifications, involving Virtual Reality and Augmented Reality respectively.
In first case a room has been created, and inside the room the three-dimensional figures have been positioned, as shown in figure \ref{fig:mesh1}.\newline
The project has then been compiled in WebGL in order to be easily used by a web browser and not have any dependence on a specific operating system (Windows, Linux, Android, iOS, etc).
As far as the use with augmented reality is concerned, Vuforia software has been used instead.
Vuforia is a framework that integrates within Unity3D.
Vuforia allows you to create projects that use augmented reality by providing all the functions essential for operation on mobile phones.\newline
In particular we focused on smartphones with Android operating system.

\section{Conclusions and future works}
In the first phase of our research we tried to understand which mathematical objects and which tools to use to enhance mathematical teaching, having sensed that the use of VR techniques significantly increase the level of understanding of the subject studied by the students.\newline
We would like to expand our research towards the field of immersive learning, in particular those applications that allow the user to be immersed in virtual worlds so that he can increase brain stimulation during the learning phase.\newline
Our approach aims to give the student a wider environment of objects to study and deepen, by selecting among them the ones of greatest interest and didactic utility.
Furthermore, we find it interesting to know the degree of interest and empathic response of students to the system: sensations, disturbances, emotions aroused; we do not intend to neglect any advice from them to make the environment more livable and attractive.\newline
That's why the next step we are considering is to allow high school students to evaluate the quality of the work done by filling out a questionnaire.\newline If the number of students is high enough, we can draw objective conclusions and understand how much virtual reality can affect their perception of mathematics.
We are therefore going to select a significant sample of them, homogeneous by social and cultural level, in order to obtain a set of coherent and indicative answers.\newline
As far as the generation of functions is concerned, the current system is able to display objects that are only statically compiled at compiling-time: for the future we intend to design a dynamic system for the generation of mathematical functions that allows the mathematics teacher to draw arbitrarily any three-dimensional or two-dimensional graph without the need to print a new Vumark every time.\newline
What we want to achieve is a dynamic platform that allows us to understand how far the freedom of choice of a function by the user, as well as a complete experience of immersion in the mathematical object itself and in its specific characteristics and properties, develops and increases in students a high level of understanding of the topics covered.

%
%
\printbibliography

\end{document}